\documentclass[sigconf]{acmart}

\usepackage[english]{babel}
\usepackage{hyphenat}
\usepackage{microtype}

\setcopyright{none}
\renewcommand\footnotetextcopyrightpermission[1]{}

\acmConference{Arxiv pre-print}{October 2024}{}

\usepackage{graphicx}
\usepackage{xcolor}

\usepackage{hyperref}
\usepackage{float}
\usepackage{tabularx}
\usepackage{booktabs}
\usepackage{balance}
\usepackage{subcaption}

\begin{document}

\sloppy

\title{An Analysis of QUIC Connection Migration in the Wild}

\author{Aurélien Buchet}
\affiliation{
    \institution{Université Catholique de Louvain, Belgium}
}
\email{aurelien.buchet@uclouvain.be}

\author{Cristel Pelsser}
\affiliation{
    \institution{Université Catholique de Louvain, Belgium}
}
\email{cristel.pelsser@uclouvain.be}

\begin{abstract}
    As QUIC gains attention, more applications that leverage its capabilities are emerging.
    These include defenses against on-path IP tracking and traffic analysis. However, the deployment of the underlying required support for connection migration remains largely unexplored.
    This paper provides a comprehensive examination of the support of the QUIC connection migration mechanism over the Internet.  
    We perform Internet-wide scans revealing that despite a rapid evolution in the deployment of QUIC on web servers, some of the most popular destinations do not support connection migration yet. 
\end{abstract}

\keywords{QUIC, Migration, Measurements}

\maketitle

\section{Introduction}
\label{intro}

Over the last decade, the QUIC protocol \cite{rfc9000} gained a lot of interest among researchers and in the industry.
The protocol is now standardized by the IETF, supported by many browsers and content providers.
The initial handshake includes TLS keys exchange such that encrypted data can be sent as soon as the handshake is completed. 
As a fast and encrypted protocol, QUIC is designed to improve the performance of the Hypertext Transfer Protocol (HTTP) \cite{rfc9114} but is also used with other protocols and applications that can benefit from its features.

In particular, QUIC's connection migration mechanism enables novel interesting use cases.
It allows hosts to switch from one IP address to another, mid-flight of a connection, while maintaining its state and avoiding the need to perform a whole new handshake.
This feature can be used to improve performance by allowing fast handover in mobile environments similar to what has been explored with Multipath TCP \cite{ford2013tcp} by Paasch et al. \cite{paasch2012exploring}.
But it also finds applications in privacy and security.
Recent research demonstrated that systems relying on QUIC and connection migration can serve as an effective defense mechanism against IP-based tracking and traffic analysis from an on-path observer \cite{govil2020mimiq,siby2023evaluating,wang2023raven}.

Despite new use cases and applications for connection migration, the adoption of these novel features by web servers remains unclear. 
Several studies already exist on the deployment of QUIC.
In 2018, before the protocol's standardization, Ruth et al. conducted a study on the number of QUIC servers and the share of QUIC traffic on the Internet \cite{ruth2018first}.
Piraux et al.~\cite{Piraux_2018} investigated the evolution of QUIC implementations as the drafts were released, testing the changes proposed by different versions of the drafts. 
Right before the release of the RFC, in 2021, Zirngibl et al. \cite{Over9000} presented various methods to identify QUIC capable targets, studied the deployed versions and common configuration parameters used by servers.
Recently, Zirngibl et al. \cite{zirngibl2024quic} attempted to classify the QUIC libraries present in the current Internet.

Unfortunately, none of the aforementioned studies investigated the support for connection migration.
Our work aims at filling this gap in the literature with new measurements focused on the deployment of this feature. 
For this study, we:
\begin{itemize}
    \item Reproduced the methodology used by Zirngibl et al. \cite{zirngibl2024quic} to identify QUIC capable targets.
    \item Introduce a new tool able to perform client-side QUIC connection migration and allow us to have a first view of its server-side support at scale.
    \item Quantify the support for connection migration and compare our results for the baseline QUIC deployment with other studies.
\end{itemize}

The rest of this paper is organized as follows. 
Section \ref{Background} presents the necessary background required to understand our work.
Section \ref{methodo} explains the methodology used to perform our experiments.
We present our results for the QUIC deployment and support for connection migration in section \ref{deployment}. In Section \ref{limitations} we then highlight a set of limitations of our approach. We conclude in section \ref{conclusion}.

\section{Background}
\label{Background}

This section briefly presents the QUIC protocol and explains its connection migration mechanism. 

\subsection{QUIC}

QUIC is a reliable and secured transport protocol built on top of UDP.
It was designed to answer some shortcomings of TCP and replace it for connection-based applications.
Data transfers are stream-based allowing a one-to-one mapping with HTTP/2 streams.
Streams can be multiplexed and data delivery of a stream is independent of packet losses affecting other streams, reducing head-of-line blocking.
They are attached to a connection and not to a specific path. In addition, a stream can be migrated to another interface without the need to retransmit the data.
QUIC uses \textit{Connection Identifiers} (CIDs) to be able to link packets from multiple interfaces to the same connection.
Each QUIC packet header contains a Destination CID (DCID) in cleartext to identify the session to which it belongs.
Handshake headers also contain a Source CID (SCID) to provide an initial CID to the other endpoint.
The IETF standardized the protocol in May 2021 and designed HTTP/3 \cite{rfc9114} to take advantage of QUIC for web services.\\

\begin{figure}[tb]
    \centering
    \includegraphics[width=.5\textwidth]{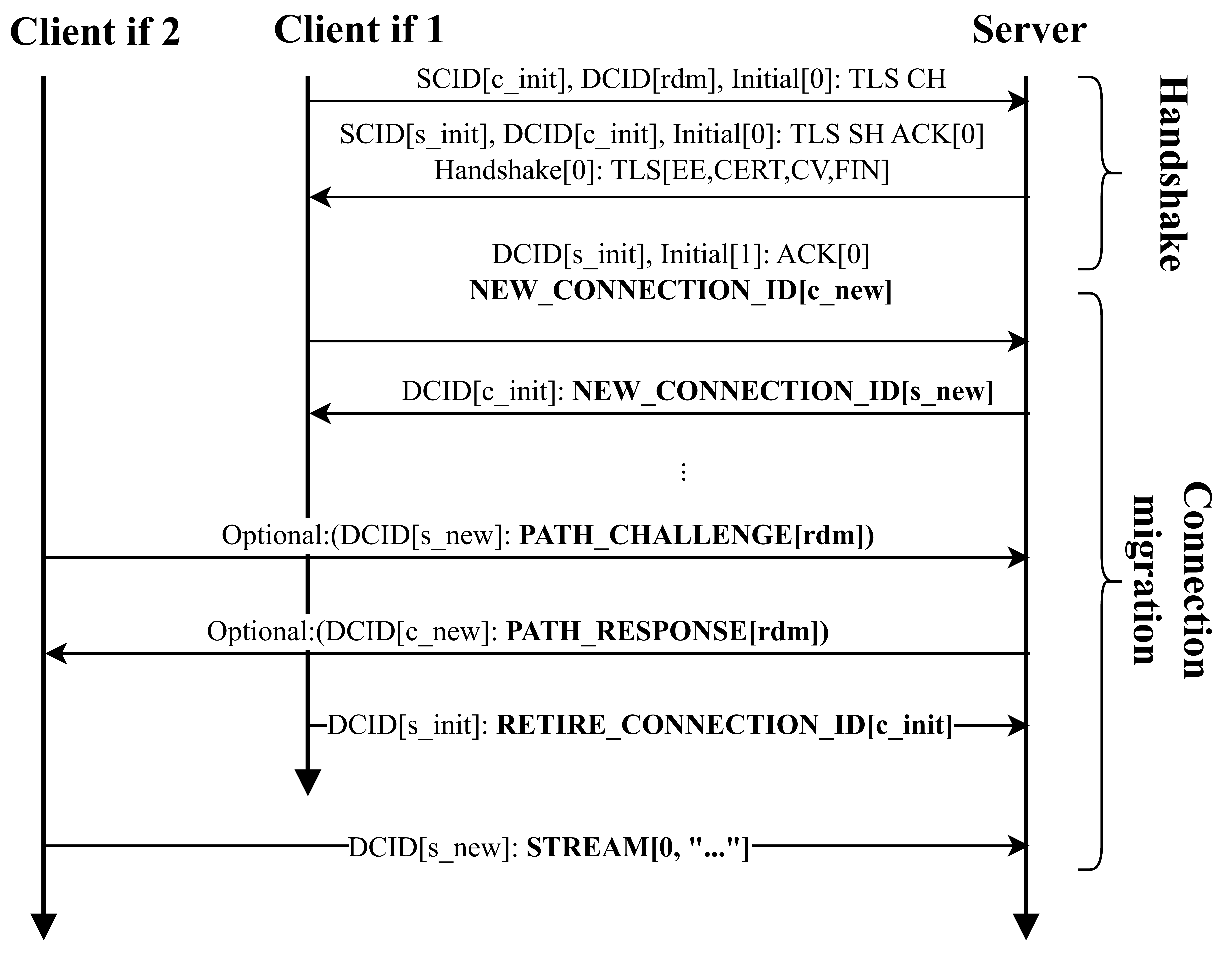}
    \caption{Example of a QUIC handshake followed by connection migration. Bold text indicates encrypted data.}
    \label{fig:quic_migration}
\end{figure}

\textbf{Handshake.} A QUIC handshake begins by sending a packet with a SCID chosen by the client and a random value for the DCID. This packet contains an \textit{Initial} frame, including a TLS 1.3 Client Hello (CH) as well as the transport parameters used, such as the QUIC version. 
To establish the connection, the server replies with its own \textit{Initial} frame containing a TLS 1.3 Server Hello (SH) in a packet with the client SCID as DCID and its own chosen CID as SCID.
\linebreak
An example of a simple QUIC handshake is shown in Figure \ref{fig:quic_migration}.

\textbf{Connection migration.} 
In order to prepare a connection migration, an endpoint can send a \textit{New\_Connection\_ID} frame to announce that another CID can be used to reach it.
These frames are encrypted and can be sent at any time during the connection.
The new CIDs can be used when a migration occurs or to resume a session after the connection is closed.
In order to render the migrated traffic unlikable to past traffic for an eavesdropper, both the source and destination CIDs need to be changed meaning that both endpoints need to have received at least one other CID.
Otherwise, the eavesdropper could link the two connections by matching the CIDs present in clear in the headers.
If there are enough CIDs on both sides of the connection, the client can attempt to migrate the connection to another interface as illustrated in Figure \ref{fig:quic_migration}.

Connection migration starts with an optional probing phase where the client sends a QUIC packet to the server using a previously negotiated CID as DCID.
This packet contains an encrypted \textit{Path Challenge} frame with a random value that the server must include in a \textit{Path Response} frame to validate the new path.
Once the new path is validated, the client can signal that the previous interface is no longer used by sending a \textit{Retire\_Connection\_ID} frame.
If the CID is not retired, it is possible to migrate back to the previous interface by using the first set of CIDs.

\section{Methods, tools and datasets}
\label{methodo}

In this section, we explain the different methods, tools and datasets used for our measurements.
All software tools and datasets used in this study are open-source and publicly available except for the migration scanner described in section \ref{migr_tool} and specific analysis scripts developed for this study.
We plan to release the source code of the latter scanner as soon as possible.

Figure \ref{fig:data_pipeline} shows the different steps in our scanning to test the support for QUIC and connection migration. The rest of the section details each step.

\subsection{Collection of IP addresses}

The approach used to find QUIC capable targets is based on techniques from previous studies \cite{Over9000, zirngibl2024quic, doq_measurements} relying on Zmap \cite{zmap}.
ZMap is a tool designed to perform large network scans. 
Its modules allow to define custom packets to send per (IP, port) pair and define a set of output values generated from the responses that are written to a csv file.
These values can include the support for a specific protocol, the version used or the RTT.
This makes it very effective to detect hosts supporting a specific protocol on a given port.
The QUIC handshake has a version negotiation mechanism that we can use to detect QUIC capable targets.
Specifically, some ranges of version numbers are reserved for version negotiation and can be used to force a version negotiation.
By sending a packet with a reserved version number and a valid Client Hello, we can detect targets responsive to QUIC handshakes.
We performed IPv4 scans targeting the whole address space on port 443 (HTTPs).

For IPv6, the massive address space makes it impossible to scan every single available address.
Instead, we relied on the responsive IPv6 addresses Hit-list service provided by Gasser et al. \cite{IPv6Hitlist}.
We also used the DNS records to find additional IPv6 addresses (left side of Figure~\ref{fig:data_pipeline}) and tested the responsive addresses using zmap on port 443.

\subsection{DNS records}

As Content Delivery Networks (CDN) grow more popular, it is quite common that multiple domains are hosted on a single server.
For this purpose, the server relies on the Server Name Indication (SNI) contained in the TLS Client Hello to direct requests to the correct website (i.e. to the right domain).
Because our Zmap scans only provide a list of IP addresses with no further indication regarding domains hosted, connections could fail to be established if no SNI is provided in our TLS Client Hello.

To address this problem, we gathered over 125 M domain names from the Centralized Zone Data Service (CZDS) and the Tranco list \cite{LePochat_2019}. We then performed DNS queries using MassDNS via a local Unbound server.
We used the A and AAAA records to create a database of IP addresses linked to domain names.
The AAAA records were also used to find additional IPv6 addresses to seed the zmap scans, as mentioned in the previous section.
We then mapped the addresses discovered during the Zmap scans to the domain names to provide SNI for the QUIC handshakes in our migration scans. 

\subsection{Connection migration tool}
\label{migr_tool}
In order to test the support of connection migration of the previously discovered targets, we use a stateful scanner capable of opening connections and migrating them.
We could not use the QScanner introduced by Zirngibl et al.
\cite{Over9000} as the underlying QUIC library, quic-go \cite{quic-go} does not support migration at the time of writing. 
We instead developed our own scanner based on Cloudflare Quiche \cite{Quiche}.
The tool is aimed specifically at testing the support for connection migration of HTTP/3 servers as it is the most common use case for QUIC at the moment.
The whole code base is around 1.2k lines of Rust code.
The scanner can be used to scan IP addresses with or without SNI.
It attempts to establish a QUIC connection and, when the handshake is successful, generates a new CID and sends it to the server.
The TLS Application-Layer Protocol Negotiation (ALPN) is used with h3 to ensure that the server supports HTTP/3.
If the server provides at least one additional CID, a migration is triggered on a provided interface by sending a \textit{Path Challenge} frame.
If the new path is validated, our scanner also sends a simple HTTP request to get the root document of the server.
This allows to retrieve additional information about the server in the HTTP headers and confirm that it is still possible to communicate with the server after the migration.

The scanner reports information about which targets are able to perform a successful handshake and migration. 
It also indicates errors that might have occurred and the HTTP server header if one was received.

We tested the scanner against a standard Cloudflare Quiche server in a controlled environment. We used different configurations to verify that the scanner was able to perform successful migrations when the server allowed it and to detect when the server did not support migration.
We also tested the scanner against a few web servers for which we could verify beforehand the support or lack of support for connection migration.
These include servers from Google and Cloudflare as they are openly deploying QUIC.

\subsection{Post-processing}
\label{postproc}
In order to identify the main actors in the QUIC deployment and the support for connection migration, we used public datasets to map the IPs to Autonomous Systems (ASes) and organizations.
We first use the routing data from RouteViews \cite{routeView} to retrieve a BGP full-feed and map prefixes to ASes.
We then obtained the name of the organizations by matching the AS numbers in the CAIDA \textit{AS to Organization Mapping Dataset}~\cite{AS2org}. Specifics on these datasets are given in section \ref{setup} for reproducibility purposes.\\

\begin{figure}[tb]
    \centering
    \includegraphics[width=.49\textwidth]{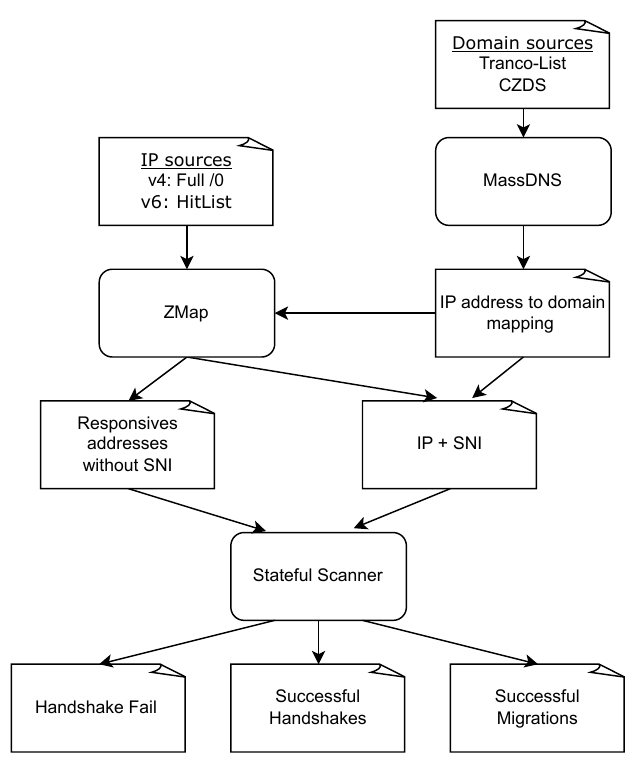}
    \caption{Methodology for the connection migration scans. Rectangles represent data and rounded rectangles represent software tools.}
    \label{fig:data_pipeline}
\end{figure}

\section{Experimental setup}
\label{setup}
We performed our experiments from two different vantage points (VP) located in two different countries.
The first vantage point, VP1, is located in a cloud provider in western Europe and the second one, VP2, in a data-center in North America. 
The deployment of a second vantage point on a different continent served as a sanity check, allowing us to verify our measurements and verify potential biases arising from local phenomena such as DNS resolvers or local firewalls.
As we did not observe any significant differences between the results from the two vantage points, we only present the results from VP1 in the main text.
The results from VP2 are available in the appendix.  

During our scans, we paid attention to follow the best practices for Internet measurements \cite{kenneally2012menlo}.
We applied a rate limit to all our scans to avoid overloading our providers and the targets.
Both vantage points host a web server with a description of the study together with a contact email address for any question or request to opt-out.
We maintained a list of prefixes that requested to be excluded from our scans and made sure to not include them in our study.

We considered the handshake to be successful only after reception of a complete and valid TLS Server Hello.
We performed the migration on the same interface but with a different source port. The migration was considered successful only if the server replied with a valid \textit{Path Response} frame.

Our study was conducted in May 2024.
We ran the whole pipeline represented in Figure \ref{fig:data_pipeline} in parallel from both vantage points.
We picked the last available version of every dataset. The different datasets that we used are: 
\begin{itemize}
    \item IPv6 Hitlist generated on the 4th of May 2024.
    \item CZDS com, net and org zones from the 10th of April 2024.
    \item Tranco list from the 25th of April 2024 \footnote{Available at \url{https://tranco-list.eu/list/24N49}}.
    \item RouteViews RIBs from the 17th of May 2024, noon. We used the Equinix collector data and picked a full feed from one IPv4 peer and one IPv6 peer.
    \item CAIDA AS to Organization Mapping Dataset from the 1st of May 2024.
\end{itemize}

\section{The state of QUIC Connection migration deployment}
\label{deployment}

Here we report on our measurements. We first present the number of responsive addresses on UDP port 443. Then, we investigate the successful handshakes and migrations on the same interface.

\begin{table*}[tb]
    \centering
    \caption{QUIC targets and connection migration support from the first vantage point. Percentages are calculated with respect to the total number of targets for handshakes and with respect to the number of successful handshakes for migrations.}
    \begin{tabularx}{.86\textwidth}{c|rrrrrr}
    \toprule
        &  \textbf{Targets} & \textbf{\begin{tabular}[c]{@{}c@{}}Distinct \\ ASes\end{tabular}} & \textbf{\begin{tabular}[c]{@{}c@{}}Successful \\ Handshakes\end{tabular}} &
        \textbf{\begin{tabular}[c]{@{}c@{}}Distinct \\ Handshakes ASes\end{tabular}} &
        \textbf{\begin{tabular}[c]{@{}c@{}}Successful \\ Migrations\end{tabular}} &
        \textbf{\begin{tabular}[c]{@{}c@{}}Distinct \\Migration ASes \end{tabular}} \\ \hline
        IPv4 no SNI & 12,024,542 & 11,854 & 591,848 (4.9\%) & 6,155 (52\%) & 11,854 (2\%) & 474 (7.7\%)\\
        IPv4 with SNI & 332,682 & 2,667 & 230,889 (69\%) & 2,241 (84\%) & 119,779 (52\%) & 1,503 (67\%)\\
        IPv6 no SNI & 3,237,165 & 3,288 & 110,764 (3.4\%) & 2,604 (79\%) & 1,281 (1.2\%) & 176 (6.8\%)\\ 
        IPv6 with SNI & 1,463,071 & 488 & 887,330 (61\%) & 426 (87\%) & 693,461 (78\%) & 227 (53\%)\\
    \bottomrule
    \end{tabularx}
    \label{targets}
\end{table*}    

\subsection{QUIC responsive targets}

The zmap scans allowed us to identify over 12~M responsive IPv4 addresses from nearly 12~k different ASes. Additionally, from the 26~M IPv6 addresses gathered, we found 3.2~M responsive addresses. The results are similar for both vantage points and on par with the latest studies \cite{zirngibl2024quic}.
These are UDP scans on port 443. A host responding on this port is likely to support QUIC. These results are consequently indicative of the number of QUIC capable HTTP servers which rapidly increased over the last few years.
The amount of responsive IPv4 addresses obtained is nearly 5 times larger than the one observed by Zirngibl et al. \cite{Over9000}, with a similar method in 2021. Moreover, the measured deployment in 2021, was already 3 times bigger than the number of addresses recorded by Rüth et al. \cite{ruth2018first} in 2018.

We were able to collect over 120~M domain names from the CZDS zones and the Tranco list.
The resolution of these domain names yielded 3~M distinct IPv6 addresses and 6.2~M distinct IPv4 addresses.
The IPv6 addresses combined with the 24.5~M addresses from the Hitlist were used to seed a zmap scan yielding 3.2~M responsive IPv6 addresses out of which we were able to map 1.4~M to at least one domain name (IPv6 with SNI targets in Table~\ref{targets}).
For IPv4, out of the 12~M responsive addresses, we were able to map 332 k to a domain name. These are the IPv4 targets with SNI in Table~\ref{targets}. 

\subsection{Stateful scans}

We performed stateful scans to test the support for HTTP/3 and connection migration.
From both vantage points, we tested the support of HTTP/3 and connection migration for the IPv4 and IPv6 targets previously identified.
We ran the scans without SNI on all the targets and with SNI on the targets for which we were able to find at least one domain name.
The results from VP1 are presented in Table \ref{targets}. The results from VP2 are similar and available in the appendix.
Each scan provides a different view of the current deployment of QUIC and the support for connection migration.

\subsubsection*{QUIC support}
The number of IPv4 targets without SNI is by far the largest as the whole address space was scanned.
It also targets the highest number of ASes.
However, the success rate for the handshake is low with only 5\% of the targets able to perform a successful handshake.
The addition of SNI from the DNS records allows to drastically increase the success rate of handshakes to almost 70\% but with a much smaller initial number of targets.
The targets are limited to domains that appear in the top lists or in the CZDS zones data which tend to be biased towards popular websites and cloud providers.
For IPv6, because we rely on Hitlist and DNS data, even the scan without SNI will have some form of bias. Indeed Steger et al. showed that the IPv6 Hitlist tends to be biased towards ISP networks \cite{steger2023target}.
The success rate for the handshake is slightly lower for IPv6 than for IPv4.
This is likely due to the fact that IPv6 addresses change more frequently than IPv4 addresses and some of the targets might have switched to a different IP address between the time the data was collected and the time of the scan.
As it is easier for providers to have a large number of IPv6 addresses, the number of addresses mapped to a domain is higher for IPv6 but the number of ASes is lower suggesting that a small number of providers are hosting a large number of domains.

The low success rate for handshakes without SNI was expected as it is required for HTTP/3 when the server is identified by a domain name.
The successful handshakes without SNI are the result of either badly configured servers or servers that are not using a domain name at all.

\subsubsection*{Connection migration support}
The migration support in IPv4 is very low for the scan without SNI with only 2\% of the targets able to perform a successful migration. 
This number is much higher for the scan with SNI reaching 52\% of successful migrations.
This indicates that the support for connection migration is available for around half of IPv4 QUIC servers discovered with an SNI.
For IPv6 on the other hand, the same trend is observed but amplified, the success rate is even lower when no SNI is provided with only 1\% of the targets able to perform a successful migration.
With SNI, the success rate is much higher with almost 80\% of the targets able to perform a successful migration after a successful handshake.
These targets are very concentrated in a small number of ASes suggesting that a few big players are supporting connection migration for IPv6.

There are a few reasons for which a server might not support connection migration.
Some implementations just do not provide support for this feature yet.
Load balancers might route the packet containing the \textit{Path Challenge} frame to a different server than the one actually handling the connection even if the server supports migration.
Firewalls might block the packets containing the \textit{Path Challenge} frame as it's a non-hanshake packet sent over a seemingly new connection.
The server might be configured to disable migration for some reason.
Unfortunately, our current version of the scanner does not record enough information to be able to determine and classify the reasons for which a migration failed.  

\subsection{Top providers}

\begin{figure*}[h]
    \centering
    \begin{subfigure}[t]{0.49\textwidth}
        \centering
        \includegraphics[width=\linewidth]{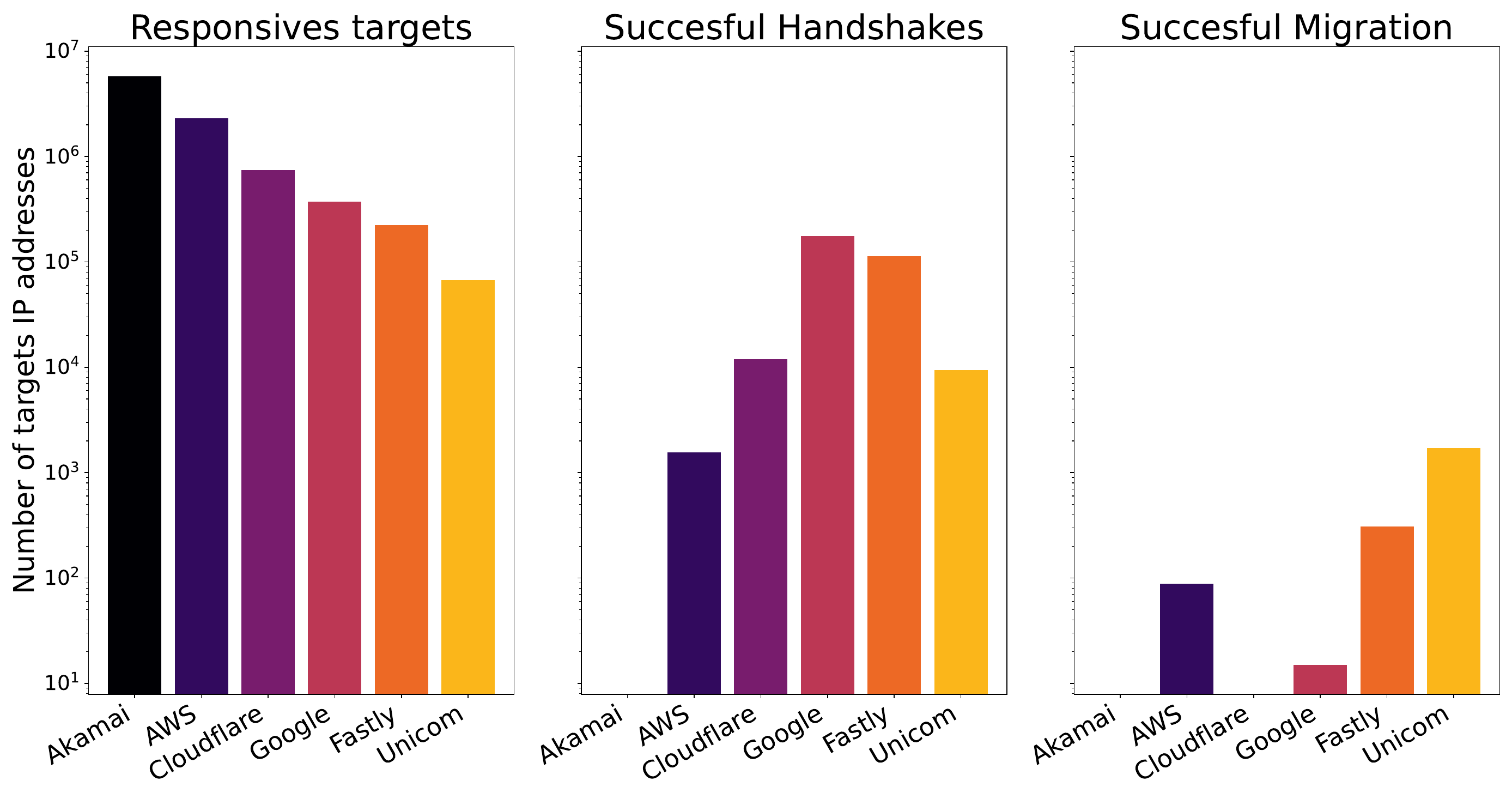}
        \caption{IPv4 no SNI}
    \end{subfigure}
    \hfill
    \begin{subfigure}[t]{0.49\textwidth}
        \centering
        \includegraphics[width=\linewidth]{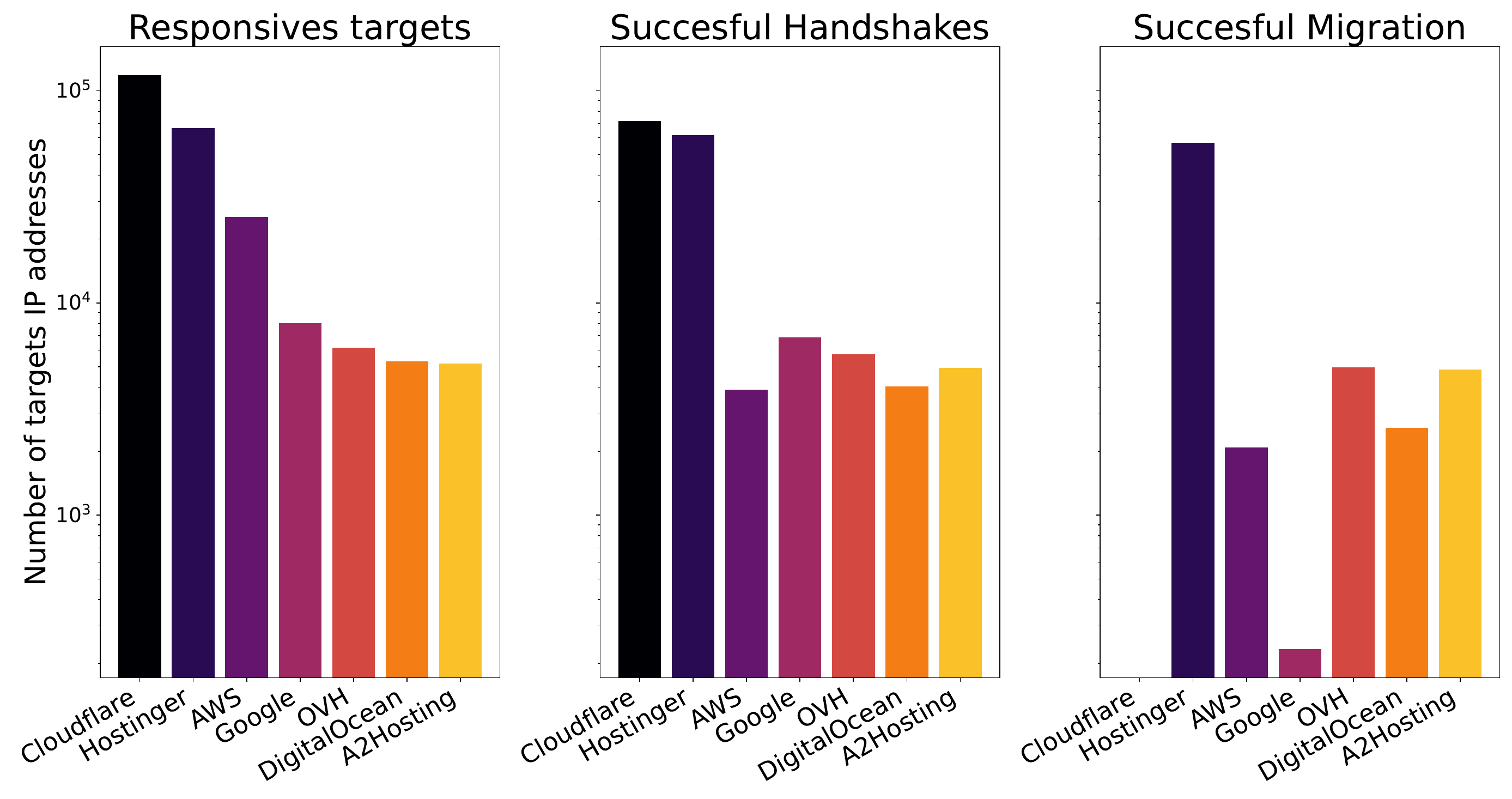}
        \caption{IPv4 with SNI}
    \end{subfigure}
    
    \begin{subfigure}[t]{0.49\textwidth}
        \centering
        \includegraphics[width=\linewidth]{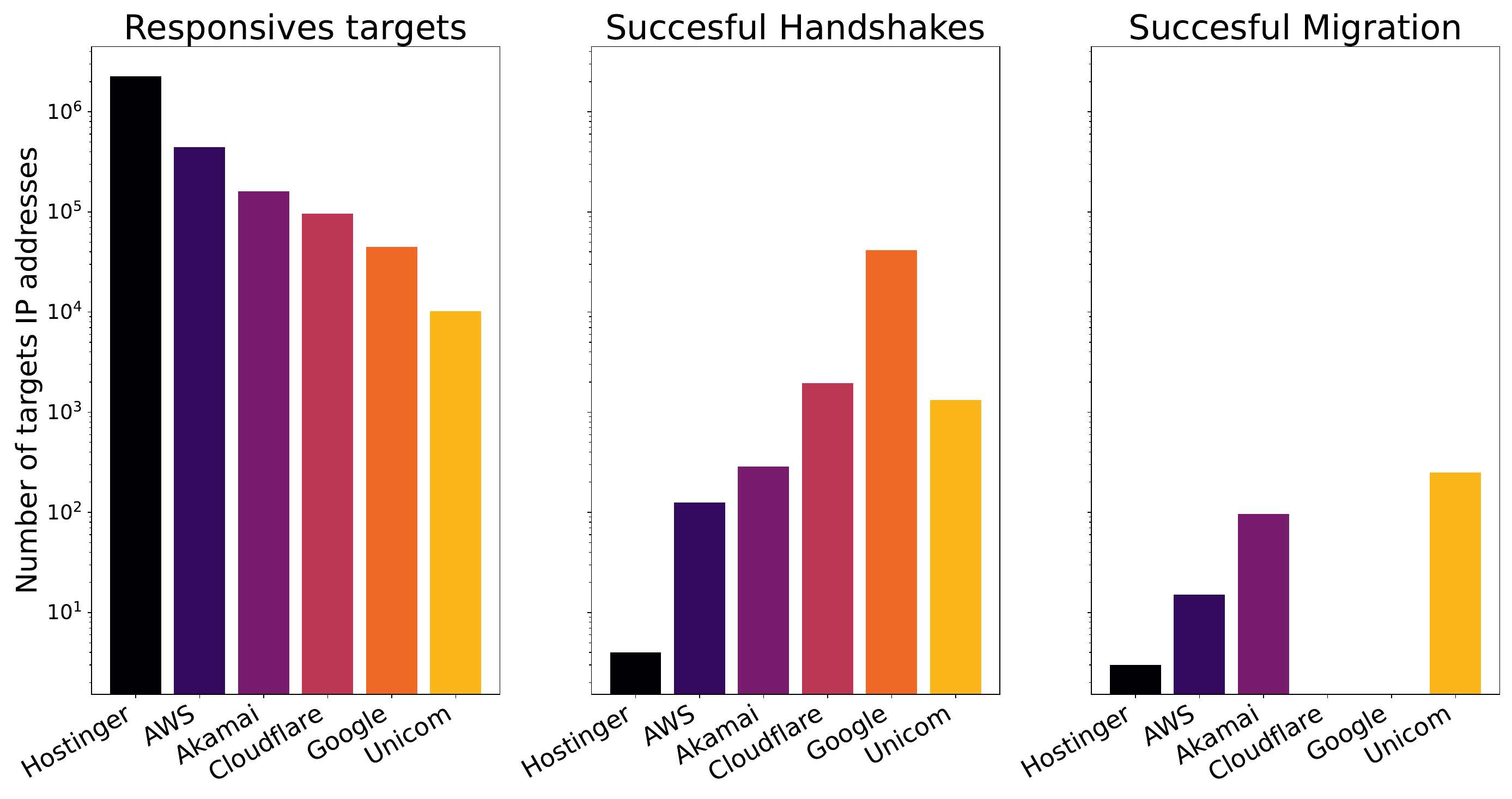}
        \caption{IPv6 no SNI}
    \end{subfigure}
    \hfill
    \begin{subfigure}[t]{0.49\textwidth}
        \centering
        \includegraphics[width=\linewidth]{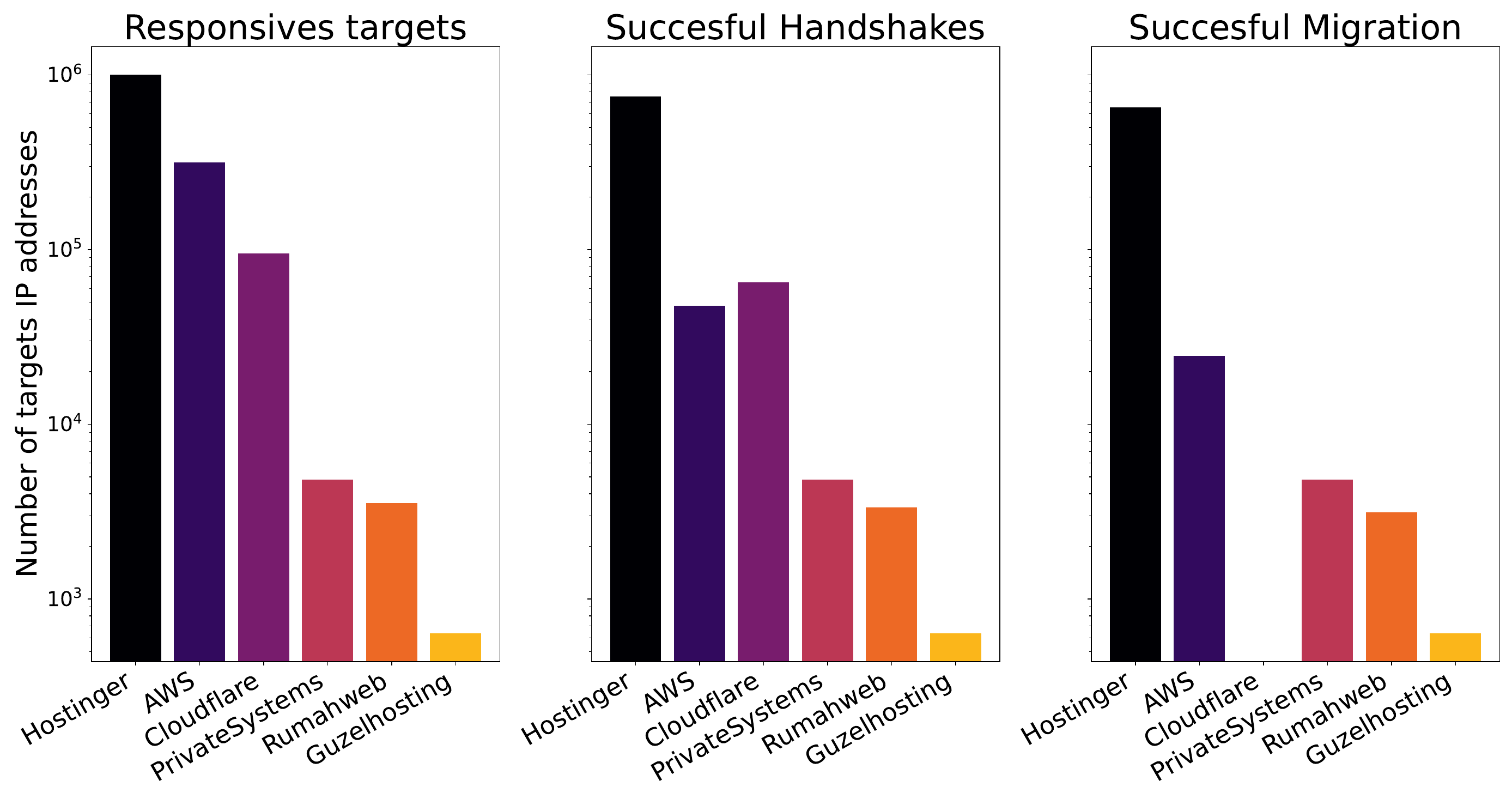}
        \caption{IPv6 with SNI}
    \end{subfigure}

    \caption{Top Providers for QUIC services }
    \label{top_providers}
    \end{figure*}

In an attempt to identify the main actors involved in our scans, we mapped the IP addresses of the targets to their organizations using the methodology described in section \ref{postproc}.
The top organizations for each scan are presented in Figure \ref{top_providers}. The scale of the y-axis is logarithmic.

The results show how the different scans highlight different actors of the QUIC deployment.
For the IPv4 scan without SNI, the QUIC responsive targets are dominated by Akamai servers which do not appear at all in the scan with SNI.
The other top providers are hyper-giant tech companies such as Amazon, Cloudflare and Google.
Among them, only Google is able to perform handshakes without SNI for most of the addresses targeted.
Cloudflare and Amazon lose several orders of magnitude when no SNI is provided while Akamai is not present at all.
The scan with SNI is dominated by Cloudflare, Hostinger and Amazon.
Interestingly, a lot of the targets from AWS are not able to perform a successful handshake despite the presence of the SNI.
Cloudflare and Google do not seem to support connection migration while the other providers do.

For the IPv6 scans, some of the observations are similar to the ones made for the IPv4 scans.
Akamai only appears in the scan without SNI and mainly Google is able to perform a successful handshake without SNI.
The most notable difference is the presence of a large number of targets from Hostinger having a large impact on the results.
Out of the 693,461 targets supporting migration in IPv6, it appears that 654,598 are from Hostinger while the other providers seem to offer very little support for the migration.

We observe that for some providers, all targets have the same behavior, either all supporting migration or none.
For others, the behavior is more diverse with some targets supporting migration and others not.
This is likely due to differences in the type of service provided by the providers.
Cloudflare and Hostinger, offer hosting services with an already configured transport layer infrastructure while others might offer infrastructure services where the users deploy their own servers such as AWS or OVH.
We don't have a clear explanation for the high amount the successful handshakes without SNI for Google and Fastly.
We can only posit that they do not strictly enforce the use of SNI or that they have a large number of servers that are not configured with a domain name.

\subsection{HTTP server headers}

When the migration is successful, a simple HTTP request is sent to the server.
If the server replies, the server header is extracted and stored.
The top 3 HTTP server headers for the targets supporting connection migration are presented in Table \ref{http_server}.

From the scan with SNI, we can see that Hostinger uses LiteSpeed as their HTTP server.
For AWS targets, the value of the HTTP server header is AmazonS3.
The results of the scan without SNI show a majority of the targets using nginx as their HTTP server. 
Some other servers such as Jakarta Server Pages (JSP), kittenx and gvs are also present but with a much lower number of targets.

\begin{table}[h]
    \centering
    \caption{Top 3 HTTP server headers for the targets supporting connection migration.}
    \begin{tabularx}{.50\textwidth}{rrrrr}
    \toprule
        \textbf{IPv4 no SNI} & \textbf{IPv4 with SNI} &
        \textbf{IPv6 no SNI} & \textbf{IPv6 SNI} \\\hline
        nginx(7.1k) & LiteSpeed(88k) & nginx(884) &  LiteSpeed(620k)\\
        JSP3(1.3k) & AmazonS3(439) & gvs(134) &   AmazonS3(6.3k)\\
        kittenx(642) & Apache(187) & JSP3(75) &  nginx(2k)\\
    \bottomrule
    \end{tabularx}
    \label{http_server}
\end{table}

\section{Limitations}
\label{limitations}

A limitation of the scanner is that it does not store information about the parameters sent by the server even those that might be relevant for migration such as \textit{disable\_active\_migration} or \textit{preferred\_address}.
It also does not record SNI that might appear in TLS certificates when the SNI is not provided in the Client Hello.
Some QUIC implementations that support connection migration might have been missed by the scanner as we did not test it against all known implementations and possible configurations.
The QUIC Interop runner \cite{seemann2020automating} could be used to test the support of migration for a wider range of implementations but at the moment, it does not fully support connection migration testing.

\section{Conclusion and future work}
\label{conclusion}
Connection migration capable protocols such as QUIC have recently been proposed as a way to improve the privacy and security of networks.
This work investigates the actual support of this new mechanism through Internet-wide scans.
We performed scans following state of the art methods allowing us to perform large scale QUIC scans.
We also developed a new tool to test the support for connection migration of QUIC servers.
Our results show that despite promising applications for privacy and security, the support for connection migration is not yet present for all big QUIC providers.
We confirm that the adoption of QUIC, on the other hand, is growing rapidly with almost 5 times more responsive targets than studies from 2022 \cite{Over9000}.
Conducting further scans in the future will allow us to track the evolution of the deployment of the connection migration support.
Trying to understand the reasons behind QUIC servers not supporting connection migration might help to improve the deployment of the feature.
Our approach is limited to migrating the client side of the connection, using another port on the same interface.
Investigating the usage of server preferred addresses and the support for migration from one server to a preferred one would be an interesting extension of this work.
With most of the hyper-giants supporting QUIC, the deployment of the protocol is likely to continue to grow in the future meaning that applications relying on connection migration will be able to reach a larger audience.

\bibliographystyle{ACM-Reference-Format}

\bibliography{biblio.bib}

\appendix

\section*{A. Results from the second vantage point}

The results from the second vantage point can be found in Table~\ref{appendix_targets}, Figure~\ref{appendix_targets} and Table~\ref{appendix_http_server}.
They are very similar to the results from the first vantage point.
As the scans were performed in parallel and relying on the same datasets, the results are expected to be similar.

\begin{table*}[ht]
    \centering
    \caption{QUIC targets and connection migration support from the second vantage point.}
    \begin{tabularx}{.86\textwidth}{c|rrrrrr}
    \toprule
        &  \textbf{Targets} & \textbf{\begin{tabular}[c]{@{}c@{}}Distinct \\ ASes\end{tabular}} & \textbf{\begin{tabular}[c]{@{}c@{}}Successful \\ Handshakes\end{tabular}} &
        \textbf{\begin{tabular}[c]{@{}c@{}}Distinct \\ Handshakes ASes\end{tabular}} &
        \textbf{\begin{tabular}[c]{@{}c@{}}Successful \\ Migrations\end{tabular}} &
        \textbf{\begin{tabular}[c]{@{}c@{}}Distinct \\Migration ASes \end{tabular}} \\ \hline
        IPv4 no SNI & 12,024,542 & 11,854 & 591,848 (4.9\%) & 6,155 (52\%) & 11,884 (2\%) & 474 (7.7\%)\\
        IPv4 with SNI & 332,188 & 2,667 & 214,753 (65\%) & 2,111 (79\%) & 100,921 (47\%) & 1,375 (65\%)\\
        IPv6 no SNI & 3,237,165 & 3,288 & 120,064 (3.7\%) & 2,643 (80\%) & 1,182 (1\%) & 175 (6.6\%)\\ 
        IPv6 with SNI & 1,463,096 & 488 & 882,530 (60\%) & 422 (86\%)& 687,480 (78\%) & 230 (54\%)\\
    \bottomrule
    \end{tabularx}
    \label{appendix_targets}
\end{table*}  

\begin{figure*}[h]
    \centering
    \begin{subfigure}[t]{0.49\textwidth}
    \centering
    \includegraphics[width=\linewidth]{images/BBTE\_IPv4NOSNI.pdf}
    \caption{IPv4 no SNI}
\end{subfigure}
\hfill
\begin{subfigure}[t]{0.49\textwidth}
    \centering
    \includegraphics[width=\linewidth]{images/BBTE\_IPv4SNI.pdf}
    \caption{IPv4 with SNI}
\end{subfigure}

\begin{subfigure}[t]{0.49\textwidth}
    \centering
    \includegraphics[width=\linewidth]{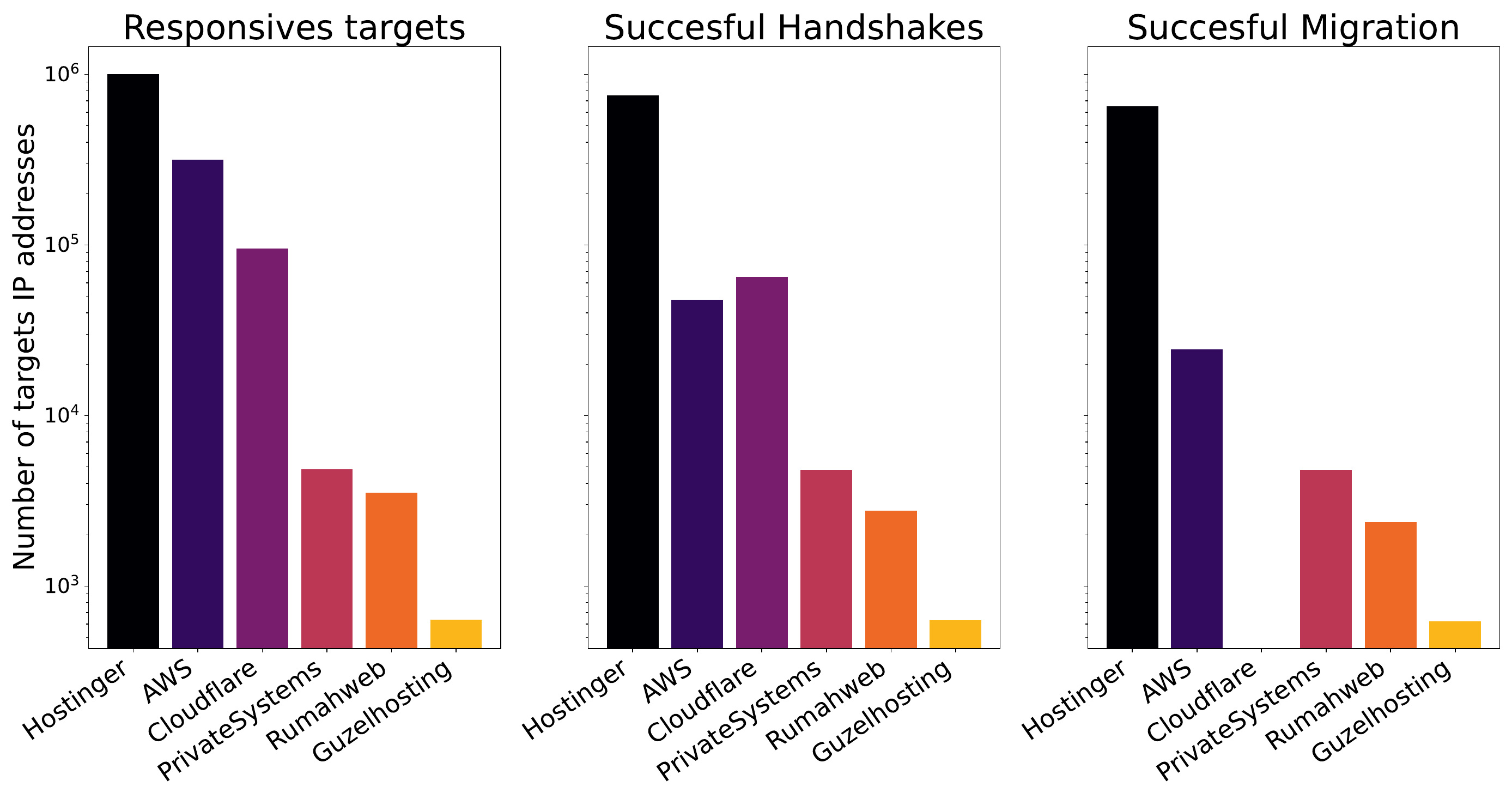}
    \caption{IPv6 no SNI}
\end{subfigure}
\hfill
\begin{subfigure}[t]{0.49\textwidth}
    \centering
    \includegraphics[width=\linewidth]{images/BBTE\_IPV6\_SNI.pdf}
    \caption{IPv6 with SNI}
\end{subfigure}
\caption{Top Providers for QUIC services on VP2.}
\label{appendix_top_providers}
\end{figure*}

\begin{table}[h]
    \centering
    \caption{Top 3 HTTP server headers for the targets supporting connection migration on VP2.}
    \begin{tabularx}{.49\textwidth}{rrrrr}
    \toprule
        \textbf{IPv4 no SNI} & \textbf{IPv4 with SNI} &
        \textbf{IPv6 no SNI} & \textbf{IPv6 SNI} \\\hline
        nginx(6.6k) & LiteSpeed(72k) & nginx(785) &  LiteSpeed(579k)\\
        JSP3(1.2k) & AmazonS3(442) & JSP3(84) &   AmazonS3(6.4k)\\
        kittenx(635) & nginx(131) & gvs(32) &  nginx(2k)\\
    \bottomrule
    \end{tabularx}
    \label{appendix_http_server}
\end{table}

\end{document}